\documentclass[aps,prd,preprintnumbers,superscriptaddress,twocolumn,nofootinbib]{revtex4-2}
\usepackage[dvips]{graphicx}
\usepackage{bm,latexsym,amsmath,amssymb,amsfonts,mathrsfs}
\usepackage{color}
\usepackage{comment}
\usepackage{here}

\def\Tr{{\text{Tr}}}

\input{colordvi.tex}

\begin{document}

\preprint{KOBE-COSMO-22-07}

\title{
Page curve and symmetries
}

\author{Pak Hang Chris Lau}
\email{phcl2@panda.kobe-u.ac.jp}
\affiliation{Department of Physics, Kobe University, Kobe 657-8501, Japan}

\author{Toshifumi Noumi}
\email{tnoumi@phys.sci.kobe-u.ac.jp}
\affiliation{Department of Physics, Kobe University, Kobe 657-8501, Japan}

\author{Yuhei Takii}
\email{yuhei.takii@stu.kobe-u.ac.jp}
\affiliation{Department of Physics, Kobe University, Kobe 657-8501, Japan}

\author{Kotaro Tamaoka}
\email{tamaoka.kotaro@nihon-u.ac.jp}
\affiliation{Department of Physics, College of Humanities and Sciences, Nihon University, Sakura-josui, Tokyo 156-8550, Japan}

\begin{abstract}

Motivated by the quantum process of black hole evaporation and its implications for symmetries, we consider a qubit system with a random dynamics as a toy model of black hole. We compute its symmetry-resolved entropies and discuss its implications. We first consider the case where charges are conserved and compute the symmetry-resolved entropies. We derive a symmetry-resolved analogue of the Page curve. We then consider the case where symmetry is explicitly broken and charges are no longer conserved. It serves as a toy model for global symmetry breaking in black hole evaporation. Despite the simple framework, the symmetry-resolved entropies capture various interesting features during the analogous process of black hole evaporation in our qubit model. 
\end{abstract}

\maketitle
\section{Introduction \label{sec:Mov}}

Thought experiments on black hole evaporation offer various lessons on symmetries in quantum gravity. The most famous one is on the absence of global symmetries, which states that symmetries in quantum gravity have to be either gauged or explicitly broken~\cite{Banks:1988yz,Banks:2010zn,Harlow:2018tng}. It is then natural to ask how large gauge interactions have to be and also how much symmetries have to be broken. Such quantitative questions are crucial when exploring phenomenological implications of quantum gravity constraints. A lot of research activities are ongoing toward this direction, especially along the line of the conjectured swampland criteria~\cite{Arkani-Hamed:2006emk,Ooguri:2006in}. See, e.g.,~\cite{Palti:2019pca,vanBeest:2021lhn,Harlow:2022gzl} for review articles.

More recently, significant progress has been made on the long-standing problem in quantum gravity, whether or not black hole evaporation follows the unitary time evolution~\cite{Hawking:1975vcx,Hawking:1976ra}, in the light of entanglement entropy and gravitational path integral~\cite{Penington:2019npb,Almheiri:2019psf,Almheiri:2019qdq,Penington:2019kki}.
Notably, this progress also made implications for global symmetries in quantum gravity~\cite{Harlow:2020bee,Chen:2020ojn,Hsin:2020mfa,Belin:2020jxr,Milekhin:2021lmq}. It is then of great interests if one can use these developments to quantify required gauge interactions or symmetry breaking. We would like to set up a framework for such a quantitative problem.

According to the progress mentioned above, the time evolution of the entanglement entropy in the evaporation process follows the so-called unitary Page curve. The original Page curve~\cite{Page:1993wv,Page:2013dx} was introduced as entanglement entropy with random unitary time evolution, assuming that the dynamics of black holes are highly chaotic. 
Such random time evolution is also used in the Hayden-Preskill protocol~\cite{Hayden:2007cs}, which gives insight into information recovery in the evaporation process. Several recent studies~\cite{https://doi.org/10.48550/arxiv.2007.00895, https://doi.org/10.48550/arxiv.2103.01876} have discussed information recovery in the Hayden-Preskill protocol when there are conserved quantities.

In this paper, inspired by these developments, we discuss a quantity called symmetry-resolved (entanglement) entropy~\cite{Laflorencie_2014,Goldstein:2017bua,Xavier:2018kqb} in similar random systems as a toy model for studying the relationship between symmetries in quantum gravity and the dynamics of black holes. See~\cite{Milekhin:2021lmq} for a seminal work in this direction in the context of two dimensional gravity.
The symmetry-resolved entropy is a finer decomposition of the entanglement entropy into superselection sectors of conserved charges. We will find that it is useful to introduce an analogue of the Page curve in the presence of symmetries. We also find that this perspective is useful to quantify effects of symmetry breaking  on entanglement entropy.

The organization of this paper is as follows. In section \ref{sec:sre}, we review the entanglement measures relevant in this paper, especially the symmetry-resolved entropy. In section \ref{sec:srpc}, we introduce our toy model of black hole evaporation with conserved charges and then derive a symmetry-resolved analog of the Page curve. In section \ref{sec:sve}, we introduce explicit symmetry breaking and study its effects on the Page curve. In section \ref{sec:dis}, we conclude by discussing the implications of our results for the evaporation process with charged Hawking radiation.


\section{Symmetry-resolved entropy \label{sec:sre}}

This section reviews the necessary entanglement measures used in this paper. Let us consider a system in a pure state represented by a density matrix $\rho= \left| \psi \right\rangle \left \langle \psi \right|$. We split the total system into a subregion $A$ and its complement $\bar{A}$. A basic entanglement measure between the two regions is the entanglement entropy. It is defined to be the von Neumann entropy of the reduced density matrix, 
\begin{align}
S_A= - \Tr \rho_A \log \rho_A \,,
\end{align}
where $\rho_A=\Tr_{\bar{A}} \, \rho$ is the reduced density matrix of the subregion $A$ constructed by tracing the total density matrix over the complement region $\bar{A}$. Entanglement entropy has the property that $S_A = S_{\bar{A}}$ for a system in a pure state. When the two regions are maximally entangled, the entanglement entropy attains its maximal value of $S_A= \min(\log d_A, \log d_{\bar{A}})$, where $d_A$ and $d_{\bar{A}}$ are the dimensions of the Hilbert space of the region $A$ and $\bar{A}$, respectively. A powerful tool to compute the entanglement entropy is the replica trick. We first compute a generalization of the entanglement entropy called the R\'enyi entropy,
\begin{align}
    S_{n} \equiv \frac{1}{1-n} \log \Tr [ \rho_A^n ] \,.
\end{align}
Then the entanglement entropy can be recovered by analytic continuation and taking the limit $S_A = \lim_{n \rightarrow 1} S_{n}$. The R\'enyi entropy is non-increasing in $n$, i.e., $S_{n_1} \geq S_{n_2}, \ \forall \ n_1 > n_2$. In this paper we utilize this property of the R\'enyi entropy and study the second R\'enyi entropy ($S_{2}$) instead of the entanglement entropy. It provides the tightest lower bound to the entanglement entropy and allows us to compute the Page curves.

Another focus of this paper is the existence of symmetry in the system which leads to conserved charges. The Hilbert space in such a system naturally splits into superselection sectors, labelled by the charge. We will explore the consequence of symmetry to the entanglement properties of the system. By assuming that the dynamics of the system is random, we can model the reduced density matrix using a probability distribution and decompose it into the form,
\begin{align}
    \rho_A &= \sum_q p(q) \rho_A(q) + \Delta \rho \,,
   \quad
    \text{Tr}_q\rho_A(q)=1\,,
    \label{eqn:crdm}
\end{align}
where $\rho_A(q)$ is the reduced density matrix in each charge sector and $p(q)$ is the probability of finding the reduced density matrix with charge $q$. $\text{Tr}_q$ is the trace over the charge $q$ subspace of $A$. We also include scenarios of explicit symmetry breaking which leads to violation of charge conservation. The term $\Delta \rho$ encodes such violation effects and allows transition between different charge sectors. We will first consider the case with $\Delta \rho = 0 $ in Section \ref{sec:srpc} and then study the case with $\Delta \rho \neq 0$ in Section \ref{sec:sve}.

The existence of conserved charges allows one to define a finer entanglement measure for each charge sector. Let us set $\Delta \rho =0$ and plug in the form of (\ref{eqn:crdm}) into the definition of entanglement entropy,
\begin{align}
    S_A &= -\Tr \rho_A \log \rho_A 
    \nonumber
    \\
    &= -\sum_q p(q) \Tr_q \rho_A(q) \log \rho_A(q) - \sum_q p(q) \log p(q)
\nonumber
\\
    &= \sum_q p(q) S_A(q) + S_{\text{c.f.}} \,,
\end{align}
where the first term contains the symmetry-resolved entanglement entropy $S_A(q) \equiv - \Tr_{q} \rho_A(q) \log \rho_A(q)$ defined as the von Neumann entropy for the reduced density matrix in the charge $q$ sector. The second term $S_{\text{c.f.}} \equiv - \sum_q p(q) \log p(q)$ is the charge fluctuation entropy~\cite{Lukin_2019}. Since each term in $S_A$ is semi-positive definite, there is an obvious relation $S_A\geq S_{\text{c.f.}}$. A symmetry-resolved R\'enyi entropy can similarly be defined and the symmetry-resolved entanglement entropy can be recovered by taking the limit $n\to1$:
\begin{align}
    S_{n}(q) &= \frac{1}{1-n} \log \Tr [ \rho_A(q)^n ] \,,\\
    S_A(q) &= \lim_{n\rightarrow 1} S_{n}(q) \,.
\end{align}

\section{Symmetry-resolved Page curves \label{sec:srpc}}

We first compute the symmetry resolved entropy in a random system with a conserved charge. Although our results are applicable to general quantum systems with finite dimensional Hilbert space, we will illustrate the calculation using a simple qubit model.

\subsection{Setup \label{sec:setup}}
We consider an $N$ qubits model with each qubit representing either a $+1$ or $-1$ charged particle. The allowed charge in this toy model is bounded by the number of qubits as $Q=-N,-N+2,\ldots, N$. To model the time evolution of a Page curve, we consider a time dependent bipartition of the system. At the initial time $t=0$, the region $A$ is empty and the whole system is in the complement region $\bar{A}$.
At each time step, we perform a random time-evolution and then one qubit of the system is selected to become part of $A$. We interpret this setup as a toy model of black hole evaporation by regarding $A$ and $\bar{A}$ as Hawking radiation and the black hole, respectively. 

More explicitly, at an intermediate time $t<N$, the state $\left| \psi \right\rangle$ is split into a subsystem $A$ with $t$ qubits and a complement subsystem $\bar{A}$ with $N-t$ qubits. For a state with charge $Q$ and a bipartition of the system into charge $q$ and charge $\bar{q}$ subregions, we employ the following notation:
\begin{align}
|\psi\rangle=\sum_{q+\bar{q}=Q}\sum_{a,\bar{a}}\frac{e^{i\delta(q_a,\bar{q}_{\bar{a}})}}{\sqrt{d(Q)}}|q_a\rangle_{A}\otimes|\bar{q}_{\bar{a}}\rangle_{\bar{A}}\,, \label{eqn:state}
\end{align}
where $|q_{a}\rangle_A$ denotes a charge $q$ state labeled by $a$ in the region $A$ and similarly for $|\bar{q}_{\bar{a}}\rangle_{\bar{A}}$. Note that the charge $q$ of the subregion $A$ takes the value $q=-t,-t+2,\ldots,t$ and similarly for $\bar{q}.$ The labels $a$ and $\bar{a}$ account for the degeneracy of the states in each charge sector. $\delta(q_a,\bar{q}_{\bar{a}})$ is a random phase whose average will be taken in the computation of entropy afterwards. $d(Q)$ is the dimension of the charge $Q$ superselection sector. More explicitly,
\begin{align}
d(Q)=\sum_{q+\bar{q}=Q}d_A(q)d_{\bar{A}}(\bar{q})\,,
\end{align}
where $d_A(q)$ is the number of charge $q$ states in $A$ and similarly for $d_{\bar{A}}(\bar{q})$. In our present qubit setup, we have
\begin{align}
&d_A(q)=\binom{t}{\frac{t+q}{2}}
\,,
\quad
d_{\bar{A}}(\bar{q})=\binom{N-t}{\frac{N-t+\bar{q}}{2}}
\,, 
\nonumber
\\
\label{d_concrete}
&d(Q)=\binom{N}{\frac{N+Q}{2}} 
\,.
\end{align}
The advantage of the notation \eqref{eqn:state} is that it can be easily modified to accommodate a more general situation, e.g., the scenario with violation of charge conservation studied in Section \ref{sec:sve}.

\subsection{Calculation of Page curve\label{sec:cal}}

We consider an initial state with a total charge $Q$. The state after a random time-evolution populates equally over all the charge $Q$ states and takes the form of (\ref{eqn:state}).

The reduced density matrix of the subregion $A$ reads
\begin{align}
\rho_A&=\text{Tr}_{\bar{A}}|\psi\rangle\langle\psi|
=\sum_q p(q)\rho_A(q)
\end{align}
with $p(q)$ and $\rho_A(q)$ given by
\begin{align}
\label{pq_symmetric}
p(q)&=\frac{d_A(q)d_{\bar{A}}(Q-q)}{d(Q)}
\,,
\\
\rho_A(q)&=\frac{1}{d_A(q)}\boldsymbol{1}_q
\nonumber\\
&\quad
+\sum_{a\neq b}\sum_{\bar{a}}\frac{e^{i(\delta(q_a,\bar{q}_{\bar{a}})-\delta(q_b,\bar{q}_{\bar{a}}))}}{d_A(q)d_{\bar{A}}(Q-q)}|q_a\rangle\langle q_b|\,.
\end{align}
Here $\boldsymbol{1}_q$ is the identity matrix in the charge $q$ subspace of $A$.

\begin{figure}
\centering
\includegraphics[width=80mm]{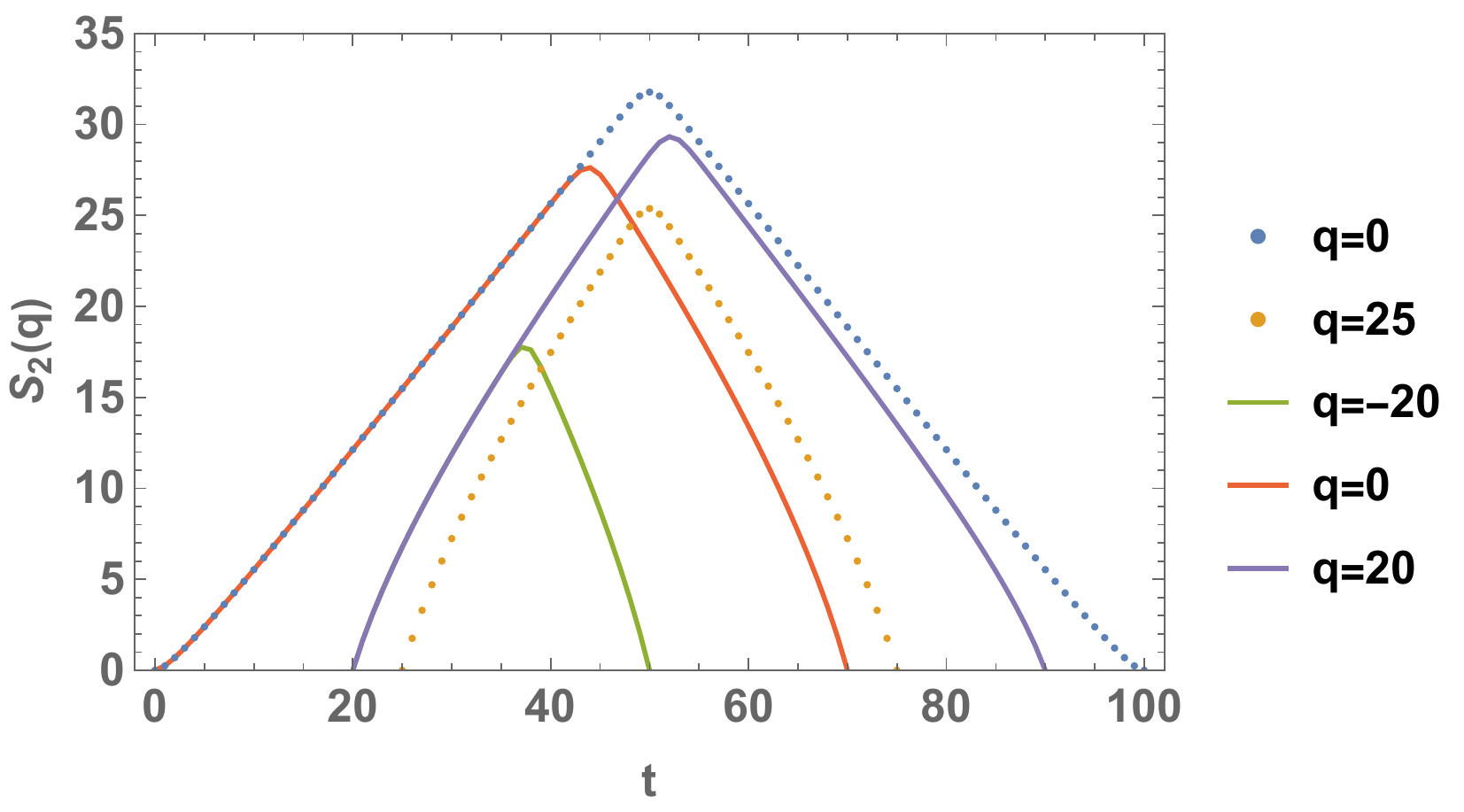}
\caption{Symmetry-resolved second R\'enyi entropy~\eqref{S_2(q)} for different values of $q$ with $Q=0$ (denoted by dotted lines) and $Q=30$ (denoted by solid lines). $N=100$ is used.}
\label{symSRrenyi}
\end{figure}

\begin{figure}
\centering
\includegraphics[width=80mm]{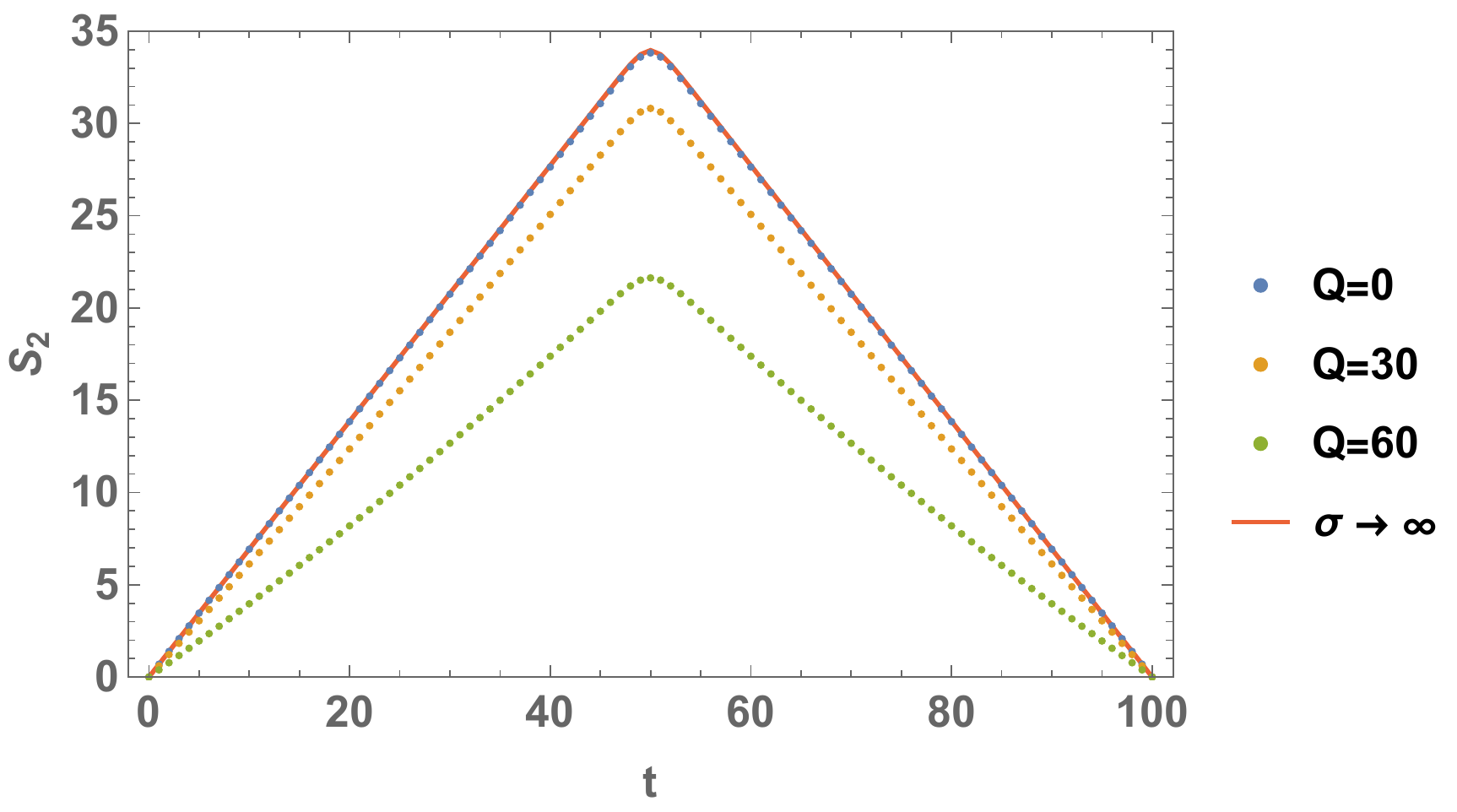}
\caption{Second R\'enyi entropy~\eqref{S_2} for the symmetry preserving case with different values of $Q$ (denoted by dotted lines) and the symmetry completely broken case, i.e., the limit $\sigma\to\infty$ of the model with the distribution function~\eqref{w_Gaussian} (denoted by a solid line). $N=100$ is used.}
\label{symrenyi}
\end{figure}

Then, the symmetry-resolved second R\'enyi entropy is
\begin{align}
e^{-S_2(q)}
&=\text{Tr}\left[\rho_A(q)^2\right]
\notag\\
\label{S_2(q)}
&=\frac{1}{d_A(q)}+\frac{1}{d_{\bar{A}}(Q-q)}-\frac{1}{d_{\bar{A}}(Q-q)d_A(q)}\,,
\end{align}
where we have taken an average over the random phases~\footnote{Such dephasing is an analog of the average over unitary operators with the Haar measure. 
To be precise, details of the sub-leading term (the third term with a minus sign) under the thermodynamic limit depend on the choice of ensemble. Refer to recent work~\cite{Freivogel:2021ivu,Stanford:2021bhl,Goto:2021mbt} that mentioned these sub-leading terms in the context of AdS/CFT correspondence. }.
In this language, the ordinary second R\'enyi entropy reads
\begin{align}
\label{S_2}
e^{-S_2}
&=\sum_qp(q)^2e^{-S_2(q)}\,.
\end{align}
Note that the expressions~\eqref{S_2(q)}-\eqref{S_2} in terms of the dimension of each Hilbert space are applicable generally, independent of fine details of the Hilbert space. 
In FIG.\ref{symSRrenyi} and FIG.\ref{symrenyi}, we plot the symmetry-resolved second R\'enyi entropy~\eqref{S_2(q)} and the second R\'enyi entropy~\eqref{S_2} in our concrete qubit model with~\eqref{d_concrete}. In particular, the former gives a symmetry-resolved analogue of the Page curve.

\subsection{Thermodynamic limit}

It is easy to derive an analytic expression of the symmetry-resolved second R\'enyi entropy~\eqref{S_2(q)} in the thermodynamic limit defined by $N\to \infty$ with $t/N$, $q/N$, and $Q/N$ fixed. For this, we introduce a symmetry-resolved analogue of the Page time $t_*(q)$ for each $q$ such that $d_A(t_*(q))=d_{\bar{A}}(N-t_*(q))$. Then, for $t<t_*(q)$, we find
\begin{align}
S_2(q)
&\simeq \log d_A(q)
\notag\\
&\simeq
\frac{(t+q)}{2}\log \frac{2t}{t+q} + \frac{(t-q)}{2}\log\frac{2t}{t-q}\,.
\end{align}
Similarly, for $t>t_*(q)$, we have
\begin{align}
S_2(q)&\simeq \log d_{\bar{A}}(Q-\bar{q})
\notag\\
 &\simeq
 \frac{(N-t+Q-q)}{2}\log \frac{2(N-t)}{N-t+Q-q}
\nonumber
\\
&\quad
+ \frac{(N-t-Q+q)}{2}\log \frac{2(N-t)}{N-t-Q+q}\,.
\end{align}

\section{Explicit symmetry breaking} \label{sec:sve}

Next we generalize the analysis to systems with explicit symmetry breaking. As we have mentioned, violation of charge conservation induces a charge-block-off-diagonal component $\Delta\rho$ of the reduced density matrix, which leads to a new contribution to the second R\'enyi entropy:
\begin{align}
\label{S2_broken}
e^{-S_2}=\text{Tr}\,\rho_{A}^2=\sum_{q}p(q)^2e^{-S_2(q)}+\text{Tr}\left(\Delta\rho^2\right)\,,
\end{align}
where the second term quantifies entanglement between different charge sectors. Our main focus here is to clarify when and how each contribution becomes dominant.

\subsection{Setup}

Let us generalize our previous setup~\eqref{eqn:state} as follows:
\begin{align}
|\psi\rangle=\sum_{q,\bar{q}}
\sqrt{w(q,\bar{q})}
\sum_{a,\bar{a}}\frac{e^{i\delta(q_a,\bar{q}_{\bar{a}})}}{\sqrt{d_A(q)d_{\bar{A}}(\bar{q})}}|q_a\rangle_A\otimes|\bar{q}_{\bar{a}}\rangle_{\bar{A}}\,,
\end{align}
where the weight function $w(q,\bar{q})\geq0$ gives a probability of finding a state in the charge $(q,\bar{q})$ sector and quantifies violation of charge conservation. It is normalized as
\begin{align}
\sum_{q,\bar{q}}w(q,\bar{q})=1\,.
\end{align}
A convenient parametrization of $w(q,\bar{q})$ for our calculations in this paper is
\begin{align}
    w(q,\bar{q}) &= \frac{d_A(q) d_{\bar{A}}(\bar{q})}{\cal N} f(q,\bar{q}) \,, \\
    {\cal N} & = \sum_{q,\bar{q}} d_A(q) d_{\bar{A}}(\bar{q}) f(q,\bar{q}) \,,
\end{align}
where ${\cal N}$ is the normalization constant and $f(q,\bar{q})$ is the distribution function controlling the degree of symmetry violation.
For example, the weight function $w(q,\bar{q})$ for the charge conserving case~\eqref{eqn:state} is localized in the subspace $q+\bar{q}=Q$ and the corresponding distribution function is
\begin{align}
f(q,\bar{q}) &= \delta_{q+\bar{q},Q} \,, \label{wsym} \\
{\cal N} &= \sum_{q} d_A(q)d_{\bar{A}}(Q-q) \equiv d(Q) \,.
\end{align}
In contrast, if the symmetry is completely broken, every state appears randomly independent of $q+\bar{q}$, so that
\begin{align}
\label{wnosym}
f(q,\bar{q}) &= 1\,, \\
{\cal N} &= \sum_{q,\bar{q}} d_A(q) d_{\bar{A}}(\bar{q}) \equiv d
\end{align}
where $d$ is the dimension of the total Hilbert space, i.e.,
\begin{align}
d=\sum_Qd(Q) \,.
\end{align}
In our qubit model $d=2^N$. We will introduce a family of weight functions interpolating the two profiles~\eqref{wsym} and~\eqref{wnosym}, but we keep $w(q,\bar{q})$ general to derive a general formula of entropy in the meantime.

\subsection{General formula}

First, the reduced density matrix reads
\begin{align}
\rho_A&=\text{Tr}_{\bar{A}}|\psi\rangle\langle\psi|
=\sum_qp(q)\rho_A(q) + \Delta \rho
\end{align}
with $p(q)$, $\rho_A(q)$ and $\Delta \rho$ given by
\begin{align}
p(q)&=\sum_{\bar{q}}w(q,\bar{q})\,,
\\
\rho_A(q)&=
\frac{\mathbf{1}_q}{d_A(q)}
\\
\notag
&\quad
+\frac{1}{p(q)}\sum_{a\neq b}\sum_{\bar{q},\,\bar{a}}\frac{w(q,\bar{q})e^{i(\delta(q_a,\bar{q}_{\bar{a}}) - \delta(q_b,\bar{q}_{\bar{a}}))}}{d_A(q)d_{\bar{A}}(\bar{q})}  |q_a\rangle\langle q_b|\,,
\\
\Delta \rho &= \sum_{q\neq q'}\sum_{a,b} \sum_{\bar{q},\bar{a}} \sqrt{\frac{w(q,\bar{q}) w(q',\bar{q})}{d_A(q)d_A(q')d_{\bar{A}}(\bar{q})^2}}
\notag\\
&\qquad\qquad\qquad
\times e^{i(\delta(q_a,\bar{q}_{\bar{a}}) - \delta(q'_b,\bar{q}_{\bar{a}}))} |q_a\rangle\langle q'_b| \,.
\end{align}
Then, the two contributions to the second R\'enyi entropy~\eqref{S2_broken} are
\begin{align}
&\sum_{q}p(q)^2e^{-S_2(q)}= \sum_{q,\bar{q},\bar{q}'} \frac{w(q,\bar{q}) w(q,\bar{q}')}{d_A(q)}
\notag\\
&\qquad\qquad\qquad\qquad
+ \sum_{q,\bar{q}} \frac{w(q,\bar{q})^2}{d_{\bar{A}}(\bar{q})}  \left[ 1-\frac{1}{d_{A}(q)} \right] \label{sbfirst}\, ,
\\
&\text{Tr}(\Delta\rho)^2 
    = \sum_{q,q',\bar{q}}  \frac{w(q,\bar{q}) w(q',\bar{q})}{d_{\bar{A}}(\bar{q})} - \sum_{q,\bar{q}} \frac{w(q,\bar{q})^2}{d_{\bar{A}}(\bar{q})}\,,  \label{sbsecond}
\end{align}
where we took an average over the random phases.
Note that the total contribution is
\begin{align}
e^{-S_2}&=\sum_{q,\bar{q},\bar{q}'} \frac{w(q,\bar{q}) w(q,\bar{q}')}{d_A(q)}+\sum_{q,q',\bar{q}}  \frac{w(q,\bar{q}) w(q',\bar{q})}{d_{\bar{A}}(\bar{q})}
\notag\\
&\quad
-\sum_{q,\bar{q}} \frac{w(q,\bar{q})^2}{d_A(q)d_{\bar{A}}(\bar{q})} \,.
\end{align}

\subsection{Illustrative examples}

\begin{figure}[t]
\centering
\includegraphics[width=80mm]{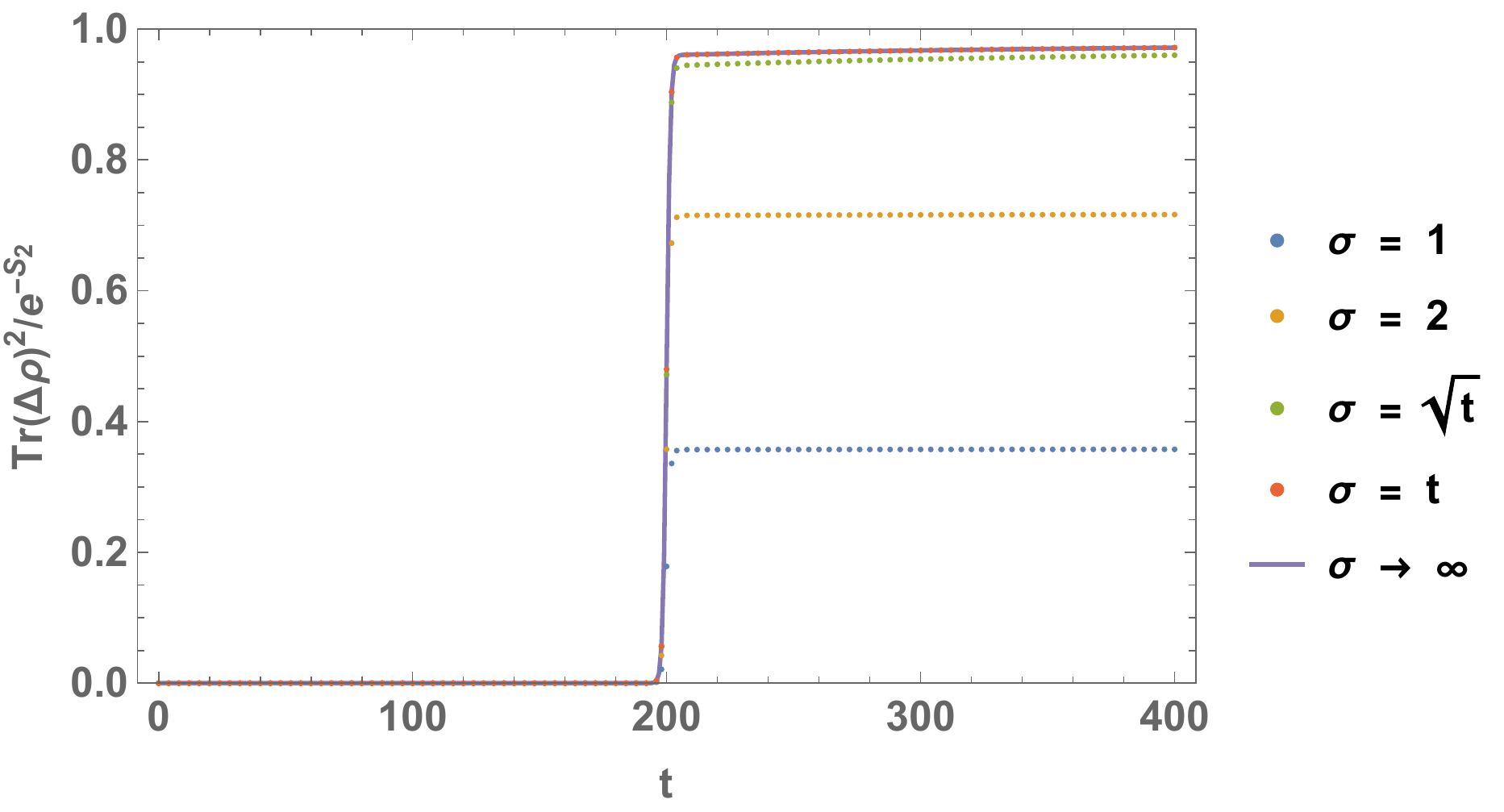}
\caption{The contribution of $\text{Tr} (\Delta\rho)^2$ to $e^{-S_2} $ for different values of $\sigma$ and $N=400$, $Q=0$.}
\label{persent}
\end{figure}

\begin{figure}[t]
\centering
\includegraphics[width=80mm]{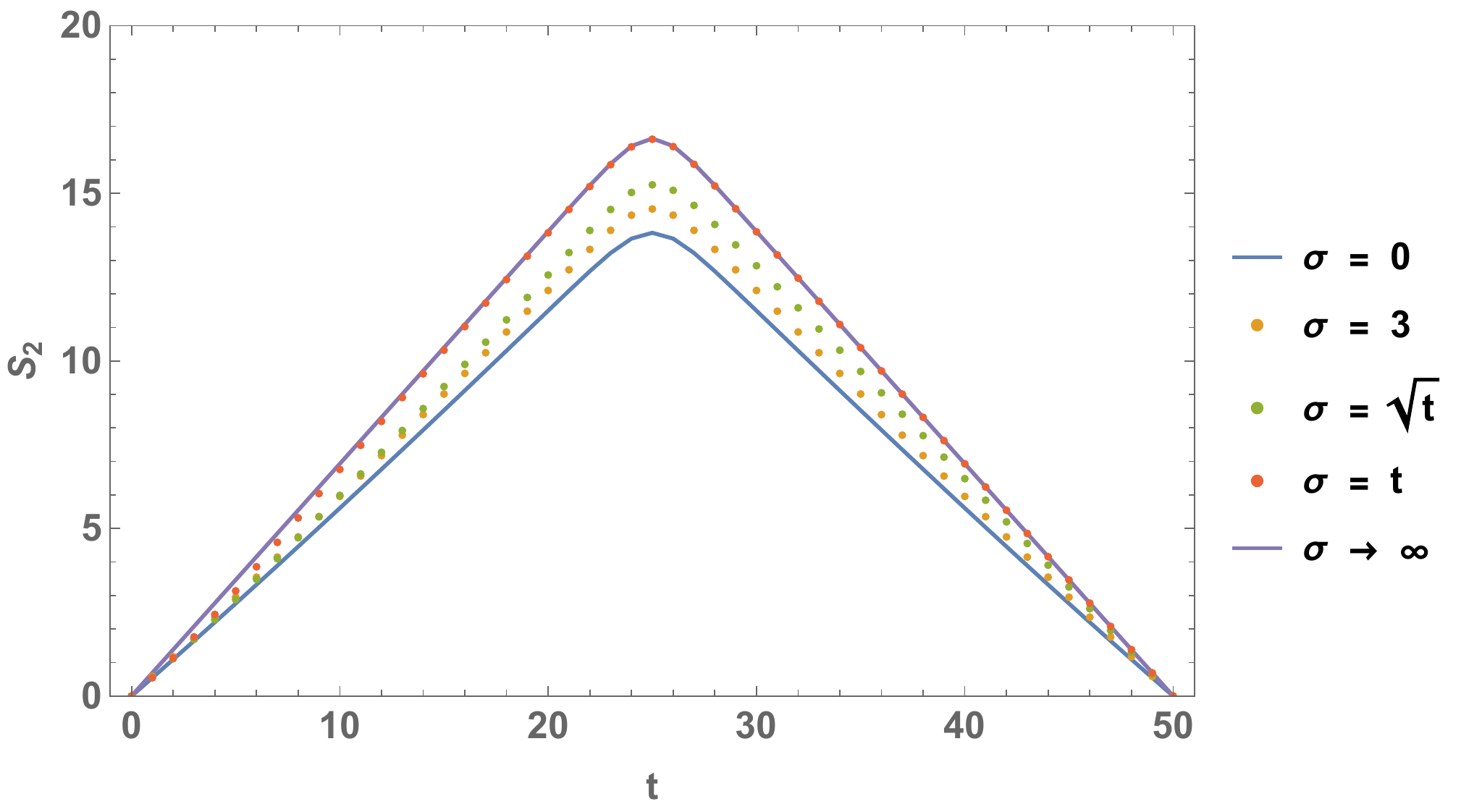}
\caption{Our results of second R\'enyi entropy $S_2$ for various values of $\sigma$ with $Q=20$. The three finite $\sigma$ cases are bounded in between by the two extreme cases ($\sigma \rightarrow 0$ and $\sigma \rightarrow \infty$). $N=50$ is used.}
\label{many}
\end{figure}

For illustration, let us consider several examples for the weight function $w(q,\bar{q})$. We begin by the extreme case~\eqref{wnosym}, where the symmetry is completely broken and time-evolution happens randomly without depending on $q+\bar{q}$. Then, the two contributions \eqref{sbfirst}-\eqref{sbsecond} read
\begin{align}
\sum_{q}p(q)^2e^{-S_2(q)}&=\frac{1}{d_A}
\left[
1+\frac{\sum_q d_A(q)^2}{d_Ad_{\bar{A}}}-\frac{1}{d_{\bar{A}}}
\right]
\, ,
\label{sbbfirst}
\\
\text{Tr}(\Delta\rho)^2 
    &=\frac{1}{d_{\bar{A}}}
    \left[1-\frac{\sum_q d_A(q)^2}{d_A^2}\right]\,,
\label{sbbsecond}
\end{align}
where we introduced
\begin{align}
    d_A=\sum_qd_A(q)\,,
    \quad
    d_{\bar{A}}=\sum_{\bar{q}}d_{\bar{A}}(\bar{q})\,.
\end{align}
We also have
\begin{align}
\label{s2bb}
e^{-S_2}=\frac{1}{d_A}+\frac{1}{d_{\bar{A}}}-\frac{1}{d_Ad_{\bar{A}}} \,.
\end{align}
Note that \eqref{sbbfirst}-\eqref{sbbsecond} and \eqref{s2bb} do not depend on the initial value $Q$ since the symmetry is completely broken.
In Fig.~\ref{persent}, we show the $t$-dependence of the contribution of \eqref{sbbsecond} to the entropy~\eqref{s2bb} with a curved line. There we find that the charge-off-diagonal contribution $\text{Tr}(\Delta\rho)^2$ becomes dominant after the Page time. The reason is actually simple: Before the Page time, diagonal components of the reduced density matrix $\rho_A$ control the entropy, so that $\Delta\rho$ does not play a role. In contrast, after the Page time, off-diagonal components become important. In models without charge conservation, the number of charge-off-diagonal components is much larger than that of off-diagonal components in the charge-diagonal sector, hence the charge-off-diagonal contribution $\text{Tr}(\Delta\rho)^2$ becomes dominant. See also FIG.~\ref{many} for comparison of $S_2$ in the present model and the symmetry preserving case studied in the previous section.

We expect that other symmetry breaking scenarios will lie in between the two extreme cases with symmetry being preserved or completely broken. We will study a particular case where the $f(q,\bar{q})$ in the weight function is chosen to be a Gaussian distribution,
\begin{align}
\label{w_Gaussian}
f(q,\bar{q})= \frac{1}{\sqrt{2\pi\sigma^2}} e^{-\frac{(q+\bar{q}-Q)^2}{2\sigma^2}} \,,
\end{align}
where the parameter $\sigma$ is the standard deviation. By taking the limit of $\sigma \rightarrow 0$, the distribution function reduces to the symmetry preserving case in \eqref{wsym} and $\Delta\rho =0$ as discussed in Section \ref{sec:srpc}. On the other hand, when $\sigma \rightarrow \infty$, symmetry is completely broken and the distribution function reduces to (\ref{wnosym}).

Now, to consider the contribution of \eqref{sbsecond} to the second R\'enyi entropy for different values of $\sigma$, we plot their $t$-dependence in FIG. \ref{persent} with dotted lines. These figures tell us that the contribution from the charge-off-diagonal component becomes significant after the Page time. As expected, the Gaussian profile~\eqref{w_Gaussian} interpolates the two extreme cases.

We end this section by comparing $S_2$ for various values of $\sigma$ with a fixed $Q$, which is a complementary calculation to FIG.~\ref{symrenyi} and the result is presented in FIG.~\ref{many}. We find that $S_2$ for a finite $\sigma$ scans a region surrounded by the symmetry completely broken case $\sigma\to\infty$ (the upper bound) and the charge conserving case $\sigma=0$ (the lower bound). Note that we chose a nonzero value of $Q$ simply to lower the peak value of the second R\'enyi entropy and make the $\sigma$-dependence clearer, but the qualitative behavior does not depend on the value of $Q$.

\section{Discussion} \label{sec:dis}

To conclude, we discuss possible implications of the results of our qubit model for black hole evaporation and symmetries. First, we consider the model with charge conservation in Section~\ref{sec:srpc}. Also, we set the total charge $Q=0$ and assume the number of qubits $N\gg 1$, having neutral black holes in mind. At the earlier time $t$ satisfying $1\ll t\ll N$, the probability~\eqref{pq_symmetric} of finding a charge $q$ state is well approximated as
\begin{align}
\label{Gaussian_charge}
p(q)\simeq\frac{1}{\sqrt{2\pi\sigma^2}}\exp\left(-\frac{q^2}{2\sigma^2}\right)\,,
\quad\sigma=\sqrt{t}\,,
\end{align}
which reproduces the Gaussian charge distribution of free Hawking particles. Note that our model did not include neutral particles such as gravitons and photons, but the generalization is straightforward and the result in the qubit model remains qualitatively the same. 

Here it is instructive to explore the consequence if we naively extrapolate the Gaussian charge distribution~\eqref{Gaussian_charge} to late time. This is analogous to extrapolating semiclassical analysis of Hawking radiation to late time of black hole evaporation. Assuming charge conservation, hence $\Delta \rho=0$, we find that
\begin{align}
\label{bound_S2}
e^{-S_2}=\sum_qp(q)^2e^{-S_2(q)}\leq\sum_qp(q)^2\sim t^{-1/2}\,.
\end{align}
Here we have neglected a $t$-independent constant that gives a subleading correction to the entropy, at the most right-hand-side. This implies that the second R\'enyi entropy monotonically increases in $t$:
\begin{align}
\label{monotinic_S2}
S_2 \geq \frac{1}{2}\log t\,,
\end{align}
which implies unitarity violation since $S_2$ is bounded by the system size as
\begin{align}
S_2\leq \min \left(\log d_A,\log d_{\bar{A}}\right)\sim \min \left( t , N-t \right)\,.
\end{align}
Note that unitarity violation in this argument happens much later than the Page time $t\sim N/2$, because the second inequality in~\eqref{bound_S2} is a loose bound (recall that we typically have $S_2(q)\gg 1$ for $t,N\gg 1$) and accordingly the monotonic growth~\eqref{monotinic_S2} is logarithmic.

This observation implies that either (A) the Gaussian charge distribution has to be modified or (B) the assumption of charge conservation has to be relaxed, in order to recover unitarity. See a recent interesting paper~\cite{Milekhin:2021lmq} for earlier discussion on this point in the context of two dimensional gravity. There, the charge distribution of the matter field (charged radiation) was essentially given by \eqref{Gaussian_charge}. Indeed, our qubit model in Section~\ref{sec:srpc} provides an example for the case (A), where the charge distribution~\eqref{pq_symmetric} is no more Gaussian, once we take into account the finite $N$ effects.

On the other hand, our qubit model with the distribution function~\eqref{wnosym} studied in Section~\ref{sec:sve} provides an illustrative example for the case (B). There, the charge distribution $p(q)$ is Gaussian simply because the symmetry is completely broken and the time-evolution is random in the charge space. However, the new contribution, $\text{Tr}\left(\Delta\rho^2\right)$ in \eqref{S2_broken}, originating from the charge-off-diagonal part $\Delta \rho$ of the reduced density matrix is responsible for decreasing the entanglement entropy to make it consistent with unitarity. We also studied a one-parameter family~\eqref{w_Gaussian} of the weight functions that interpolates the two extreme cases.

We believe that our analysis in the qubit toy models offers insights on black hole evaporation and symmetries. We hope to revisit this issue from the gravity side and report our progress in near future.

\medskip
\noindent
{\bf Note added:} After we completed the project, Ref.~\cite{Murciano:2022lsw} appeared on arXiv, which has some overlap with our Section~\ref{sec:srpc}. In Ref.~\cite{Murciano:2022lsw}, they studied the symmetry-resolved entanglement entropy and introduced a symmetry-resolved analogue of the Page curve in qubit models similar to our model in Section~\ref{sec:srpc} with charge conservation. In particular, our (16) and their (33) are consistent with each other.

\medskip
\begin{acknowledgments}

We would like to thank the organizers and participants of the Kagoshima Workshop on Quantum Aspects of Gravitation, where this project was initiated.
P.H.C.L. is supported in part by JSPS KAKENHI Grant No.~20H01902.
T.N. is supported in part by JSPS KAKENHI Grant No.~20H01902 and No.~22H01220, and MEXT KAKENHI Grant No.~21H00075, No.~21H05184 and No.~21H05462. K.T. is supported in part by JSPS KAKENHI Grant No.~21K13920 and MEXT KAKENHI Grant No.~22H05265.
\end{acknowledgments}

\bibliography{srEE}
\end{document}